\documentclass[letter]{aa}

\usepackage{amsmath}
\usepackage{graphicx}
\usepackage{natbib}
\usepackage{placeins}
\usepackage[flushleft]{threeparttable}
\usepackage{txfonts}
\usepackage{upgreek}
\bibpunct{(}{)}{;}{a}{}{,}

\begin{document}

   \title{The role of AGN jets in the reionization epoch}

   \author{V. Bosch-Ramon}

   \institute{Departament de Física Quàntica i Astrofísica, Institut de Ciències del Cosmos (ICCUB),
                          Universitat de Barcelona (IEEC-UB), Martí i Franquès 1, 08028 Barcelona, Spain \\
              \email{vbosch@fqa.ub.edu}}

   \date{Received -/Accepted -}

  \abstract %5 {} token are mandatory
  % context heading (optional)  
   {The reionization of the Universe ends the dark ages that started after the recombination era. In the case of H, reionization finishes around $z\sim 6$. Faint star-forming galaxies are the best candidate sources of the H-ionizing radiation, although active galactic nuclei may have also contributed. We have explored whether the termination regions of the jets from active galactic nuclei may have contributed significantly to the ionization of H in the late reionization epoch, around $z\sim 6-7$. We assumed that, as it has been proposed, active galactic nuclei at $z\sim 6$ may have presented a high jet fraction, accretion rate, and duty cycle, and that non-thermal electrons contribute significantly to the pressure of jet termination regions. Empirical black-hole mass functions were adopted to characterize the population of active galactic nuclei. From all this, estimates were derived for the isotropic H-ionizing 
radiation produced in the jet termination regions, at $z\sim 6$, through inverse Compton scattering off CMB photons. We find that the termination regions of the jets of active galactic nuclei may have radiated most of their energy in the form of H-ionizing radiation at $z\sim 6$. For typical black-hole mass functions at that redshift, under the considered conditions (long-lasting, common, and very active galactic nuclei with jets), the contribution of these jets to maintain (and possibly enhance) the ionization of H may have been non-negligible. We conclude that the termination regions of jets from active galactic nuclei could have had a significant role in the reionization of the Universe at $z\gtrsim 6$.}

   \keywords{Galaxies: jets - Radiation mechanisms: non-thermal - dark ages, reionization, first stars - cosmology: miscellaneous - intergalactic medium}
   \maketitle

%####################################################

\section{Introduction} 

The reionization of the Universe put an end to the dark ages that followed the recombination era, after the big bang. The reionization epoch took place mostly around $z\sim 6-8$ \citep[e.g.][]{madau17}, although the nature of the dominant H-ionizing source(s) is still uncertain. Nowadays the favoured candidates are faint star-forming galaxies, in particular the fainter and less massive ones, so long as the escape fraction of the ionizing radiation is not too low \citep[e.g.][]{stark16}\footnote{Nevertheless, it has been proposed that brighter galaxies may have dominated if the reionization duration was relatively short, and their escape fraction high enough \citep[e.g.][]{sharma18}.}. 

Another possible candidate ionizing source at high redshift is accretion in active galactic nuclei (AGN) \citep[e.g.][]{arons70,donahue87,shapiro87,meiksin93}. 
Such sources are presently considered to be a minor player in the reionization of H \citep[e.g.][]{hopkins07,onoue17,parsa18,mcgreer18}, although they are still sometimes discussed as potentially important reionization sources \citep[e.g.][]{giallongo15,grazian18}, or as indirect factors in the reionization process \citep[e.g.][]{seiler18,trebitsch18,kakiichi18}. In addition, a so far undetected, faint AGN population, as for instance accreting intermediate mass black holes in the centre of gas-rich dwarf galaxies, cannot be discarded as significant reionization sources \citep{silk17}. 

Accretion radiation is not the only form of energy output in AGN. In particular, jets may have actually dominated the energy output of AGN at $z\sim 6$: (i) the jetted AGN fraction may have been close to one (at least for the most massive black-hole AGN - e.g. \citealt{sbarrato15}-, in principle valid as well for less massive black-hole sources); and (ii) the AGN central black holes may have been accreting close to the Eddington limit with a high duty cycle \citep[e.g.][]{willott10,shankar13}.
Since jets may be as powerful as accretion radiation, or even more \citep[e.g.][]{ghisellini14,sbarrato16}, their role in the reionization of the Universe should be considered: if a substantial fraction of the jet energy went to H-ionizing photons, they may have contributed as much as AGN accretion, and perhaps even more. 

It is worth mentioning that very energetic cosmic rays may also be efficiently accelerated in AGN jets, and the impact of these cosmic rays on the ionization and heating of the intergalactic medium (IGM) at very high redshift deserves detailed studies \citep[see, in the context of POPIII microquasars, e.g.][]{tueros14,douna18,romero18}. However, cosmic ray production in AGN jets is uncertain. First, only the most energetic cosmic rays can diffuse out of the radio lobes; otherwise, they stay in the lobes and cool through adiabatic losses. Secondly, it is not known which fraction of the jet energy is in the form of very energetic cosmic rays. On the contrary, the termination region of jets are known to be filled by non-thermal electrons that could dominate pressure \citep[e.g.][]{croston18}, and these electrons could have efficiently emitted H-ionizing photons via inverse Compton (IC) scattering off CMB photons at high redshift. As bulk velocities in the lobes are relatively low, this emission would have been rather isotropic. Therefore, in this work we have focussed on the AGN jet termination regions, in particular the lobes inflated by shocked jet material, as sources of H-ionizing photons in the late phase of the reionization epoch. The beamed emission from the (relativistic) jet smaller scales is not considered here, although it could be included if it were very efficient turning jet energy into hard photons. This work is a first step in exploring the role of AGN jets and their termination regions in the late reionization epoch, say at $z\sim 6-7$; more accurate predictions are left for future studies.

The paper is structured as follows: In Sect.~2, we introduce the prescriptions adopted for the black-hole population at high redshift, together with the assumptions adopted to derive the black-hole jet power. In Sect.~\ref{jt}, we show that the termination regions of the jets of AGN at $z\gtrsim 6$ could have been powerful sources of (isotropic) H-ionizing radiation, and estimate how much this radiation may have contributed to (maintain) the reionization of the Universe. Finally, we discuss our results and conclude with a summary in Sects.~\ref{dis} and \ref{sum}, respectively.

\section{Jet-power function} 

To assess the importance of AGN jets as H-ionizing sources at $z\sim 6-7$, that is, the late reionization epoch, first one has to characterize the (total) source jet power function. This is quite difficult at very high redshifts because of the uncertainties concerning: (i) the AGN population at low  luminosities (of accretion origin); (ii) the jetted AGN fraction; (iii) the AGN (accretion) luminosity-jet power relation; and (iv) the AGN duty cycle, meaning that the fraction of the time the source was active until $z\sim 6$. One can otherwise choose to adopt a prescription for the black-hole mass function, which requires assuming, in addition to points similar to (ii), (iii), and (iv) above: (a) the yet uncertain IMBH range of the black-hole mass function \citep[say $M\sim 10^5-10^7$; see, e.g.][for a review on IMBH]{mezcua17}; and (b) the accretion-to-Eddington power ratio. We note that the empirical black-hole mass functions may differ significantly from the real ones due to incompleteness, obscuration, and inactivity (or reduced activity) \citep[e.g.][]{salvador17}. 

Here, we have adopted empirical black-hole mass functions to estimate the jet power distribution at  $z\gtrsim 6$ (we simplify the black-hole mass function in the range $z\sim 6-7$ to that for $z=6$). Two cases were adopted: the black-hole empirical mass functions at $z=6$ from \cite{shankar09} and \cite{willott10}, which reach with large uncertainties down to a black-hole mass of $M_{\rm BH}=10^5$~M$_\odot$. 
%For illustrative purposes, the corresponding black-hole, mass function (comoving) densities are shown in Fig.~\ref{bhmf}. 
Our characterization of the black hole population is similar to that of \cite{saxena17}, although in that work the authors assumed a small jetted black-hole fraction of 1\%, whereas here we consider that at $z\gtrsim 6$ most of the black holes traced by the mass functions were actively accreting and produced jets. 

To estimate the typical AGN duty cycle and accretion rate at $z\sim 6$, we followed the results by \cite{willott10,shankar13}, which point to accretion running close to the Eddington rate with a high duty cycle. In addition, motivated by hints of a dominant jetted AGN population at $z\gtrsim 4$ \citep[e.g.][]{sbarrato15}, we expected a jetted AGN fraction close to one. Finally, we assumed that the (total) jet power of one source scales with the black-hole Eddington luminosity, that is, $L_{\rm j}\propto L_{\rm Edd}$. In principle, $L_{\rm j}$ can be higher than the accretion radiation luminosity \citep[see, e.g.][]{ghisellini14,sbarrato16}.

%\begin{figure}
%       \centering
%       \vspace{-0.5cm}
%       \includegraphics[width=\linewidth]{bhmf.png}
%       \vspace{-1cm}
%       \caption{Black-hole, mass function (comoving) densities at $z\approx 6$ from \cite{shankar09} (solid line) and \cite{willott10} (dashed line).}
%       \label{bhmf}
%\end{figure}

\section{H-ionizing emission from AGN jet lobes}\label{jt}

The jets of AGN propagate through the surrounding medium. At some point, the advance of the jet head is significantly slowed down, and the jet flow is shocked, inflating a hot diluted bubble (the lobe). This bubble pushes a shocked shell of external matter, and the evolution of the whole structure is ruled by the properties of the jet, the medium, and the source age \citep[e.g.][and references therein]{scheuer74,kaiser07,hardcastle18}. In the recent Universe, two types of source are possible, depending on the morphology of the structures formed by their jets after termination \citep{fanaroff74}: Fanaroff-Riley~I (FRI), in which the jets get disrupted closer to the host galaxy and form an irregular termination region of $\lesssim 100$~kpc; and Fanaroff-Riley~II (FRII), in which the jets reach the point at which their ram pressures are in equilibrium with those of the (shocked) IGM, on scales of $\sim 1$~Mpc. 

\subsection{IC cooling and lobe dynamics}

Regardless of the AGN type - FRI or FRII - the jet inflated lobes and the lobe-shocked IGM contain together, in the adiabatic phase of the structure growth, all the momentum and energy deposited by the jet. In the radiative phase, the energy is lost through different mechanisms, thermal and non-thermal. Regarding the mass within the lobe, there could be a contribution of entrained external medium. 

When the jet-blown structure is old enough, losses due to IC on CMB photons can become relevant for its dynamics. Under equal jet and medium properties, this phase starts earlier at higher redshifts \citep[e.g.][and references therein]{saxena17}. A typical CMB photon at $z=6$, of energy $\epsilon_0\approx 3\,k\,T_{\rm CMB,z=6}\approx 5\times 10^{-3}$~eV, is IC scattered up to $\epsilon=13.6$~eV by electrons of Lorentz factor $\gamma_{\rm ion}\approx (3\epsilon/4\epsilon_0)^{1/2}\sim 50$.  
For the CMB energy density at that redshift, $u_{\rm CMB,z=6}=4\sigma T_{\rm CMB,z=6}^4/c\approx 10^{-9}$~erg~cm$^{-3}$, the IC cooling timescale is 
$t_{\rm IC}\approx 20(\gamma/50)^{-1}$~Myr (shorter at larger $z$), which is probably much less than the age of an AGN jet at $z\gtrsim 6$ with a high duty cycle (i.e. long-living). 

The larger the contribution of non-thermal electrons to the lobe pressure is, the stronger the IC impact on the lobe dynamics will be. Despite the non-thermal electron component of the pressure in these lobes is not well constrained, radio and X-ray observations show that this may easily be a substantial fraction of the total lobe pressure \citep[e.g.][]{croston18}. In addition, given the lower adiabatic index of a relativistic gas\footnote{For an ideal mono-atomic gas.} ($4/3$) with respect to that of a non-relativistic gas ($5/3$), the relativistic-to-non-relativistic pressure ratio goes as $P_{\rm rel}/P_{\rm non-rel}\sim V^{1/3}$ ($V$ being lobe volume), meaning that even moderate non-thermal pressures can eventually dominate for old enough sources. 

One can estimate the distance from the host galaxy at which the lobe isotropic IC radiation can be released. The jet propagation depends only on the medium density as long as the former is supersonic. Thus, the characteristic size of the lobe can be calculated while the jet is active as \cite[e.g.][]{kaiser07,saxena17}:
\begin{equation}
D\sim C\left(L_{\rm j}/2n_0m_{\rm a}d_{\rm gal}^\beta\right)^{1/(5-\beta)}t^{3/(5-\beta)}\,,
\label{djt}
\end{equation}
where $L_{\rm j}/2$ is used because there are two jets per source; $C$ is a constant of O(1) (set to 1 from now on); $n_0=10^{-25}/m_{\rm a}$~g~cm$^{-3}$ the medium number density at the smallest relevant spatial scale: $d_{\rm gal}\approx 6.2\times 10^{21}\,[(1+z)/3]^{-1.25}$~cm for $z>2$ (that of the host galaxy); $\beta$ the density profile power-law index ($n(r)=n_0\,(d_{\rm gal}/r)^\beta$); and $m_{\rm a}\approx 2\times 10^{-24}$~g the average particle mass. 

Taking into account that the (proper) average density of the IGM, $n_{\rm IGM}\approx 2\times 10^{-7}(1+z)^3$~cm$^{-3}$, is reached at a (proper) distance from the host galaxy of $d_{\rm IGM}\approx d_{\rm gal}(n_0/n_{\rm IGM})^{1/\beta}$ \citep{saxena17},
for a typical profile $\beta=2$ one obtains $d_{\rm IGM}\sim 20$~kpc at $z=6$. However, for $\beta=2$ the jet reaches $d_{\rm IGM}$ after a time:
\begin{equation}
t_{\rm IGM}\sim d_{\rm IGM}/\left(L_{\rm j}/2n_0m_{\rm a}d_{\rm gal}^2\right)^{1/3}\approx 4\left(L_{\rm j}/10^{44}{\rm erg~s}^{-1}\right)^{-1/3}\,{\rm Myr}. 
\end{equation}
As $t_{\rm IGM}$ is much less than the age of an AGN jet with high duty cycle, one can set $\beta=0$ for most of the source life, rendering: 
\begin{equation}
D\sim \left(L_{\rm j}/2n_{\rm IGM}m_{\rm a}\right)^{1/5}t^{3/5}\approx 30\left(L_{\rm j}/10^{44}{\rm erg~s}^{-1}\right)^{1/5}t_{\rm 10Myr}^{3/5}\,{\rm kpc}. 
\end{equation}
Taking $L_{\rm Edd}\sim L_{\rm j}\sim 10^{44}$~erg~s$^{-1}$ translates into a black-hole mass of $\sim 10^6$~M$_\odot$, although we note that the $L_{\rm j}$-dependence of $t_{\rm IGM}$ and $D$ is rather weak. 

Due to IC losses, a lower limit can be obtained for the jet length simply setting in Eq.~(\ref{djt}) $t=t_{\rm IC,bulk}$, where $t_{\rm IC,bulk}$ is the IC cooling timescale of the energetically dominant non-thermal electrons. This quantity is not well constrained, but unless these electrons have a low-energy cutoff around $\gamma_{\rm min}\gtrsim$~300, it seems likely that $t_{\rm IC,bulk}>t_{\rm IGM}$. Therefore, it is reasonable to assume that AGN jet lobes at $z\sim 6$, even those whose dynamics were dominated by IC losses, should have reached $D\gtrsim 10$~kpc \citep[see, e.g.][for similar, more accurate estimates on the lobe sizes]{saxena17}. Ionizing photons isotropically produced at those distances from the host galaxy should have easily reached the surrounding IGM, so the escape fraction was one.

It is worth mentioning that even if for some reason IC cooling had been dynamically irrelevant at $z\sim 6-7$, thermal cooling of the shocked IGM shell could have still released a significant fraction of the jet energy as UV and soft X-rays: $\sim 10$\% for a long-lasting jet with $L_{\rm j}\sim 10^{44}$~erg~s$^{-1}$, under primordial abundances \citep{sutherland93}, and reaching much farther from the host galaxy ($D>100$~kpc).

\subsection{Ionizing IC radiation}\label{ionic}

From the discussion above, it seems likely that AGN jet lobes at $z\sim 6-7$ were continuously refilled with relativistic electrons, which cooled via IC with CMB photons producing large amounts of ionizing photons. However, the actual fraction of IC scattered photons at the most suitable H-ionizing energies, say far UV, depends on the electron energy distribution. 

The electron energy distribution ($N$, where [$N$]=energy$^{-1}$) in jet lobes is expected to be cooled and
$N\propto \gamma^{-3}$ or similar, which leads to flat spectral energy distributions (SED) of synchrotron and IC radiation (i.e. $\nu L_\nu\propto\nu^0$ -or a spectral index of -1: $L_\nu\propto \nu^{-1}$-). Thus, unless $\gamma_{\rm min}>\gamma_{\rm ion}$, or $N$ is much steeper than $\propto \gamma^{-3}$ and $\gamma_{\rm min}\sim 1$ (meaning less energy at $\sim \gamma_{\rm ion}$), a substantial fraction of the jet energy should have been released in the UV through IC. 
Even if electrons are injected with $\gamma_{\rm inj}>\gamma_{\rm ion}$, a $\gamma_{\rm min}>\gamma_{\rm ion}$ is not expected under fast IC losses in the Thomson regime, as $N$ below $\gamma_{\rm inj}$ becomes $\propto \gamma^{-2}$ down to the minimum cooled electron energy (i.e. $\gamma_{\rm min}$). Such $N$ below $\gamma_{\rm inj}$ implies a moderate hardening of the SED (i.e. $\nu L_\nu\propto\nu^{0.5}$), but much softer than found, for example, by \cite{wu17} at $\gamma\sim 100$ for two radio galaxies at $z\sim 4$. The hard electron distribution of those two objects at $\gamma\sim 100$ may be explained, among other factors, by less efficient IC cooling, as $z$ is lower. 
 
We conclude that AGN jet lobes at $z\sim 6$ may have radiated through IC emission a substantial fraction of the jet energy as H-ionizing photons. The synchrotron (radio) emission from the lobes, produced by electrons with $\gamma\sim 10^4\gg \gamma_{\rm ion}$ (for magnetic fields not very far below equipartition), is expected to be well below the IC emission due to the cooling dominance of IC in Thomson. On the other hand, the synchrotron (radio) emission from the jet termination shocks (the hot spots), may have been as bright as the lobe IC emission because of a higher local magnetic field \citep{wu17}. 

Taking a cooled population of electrons with $N\propto \gamma^{-3}$, one can approximate the flux for one source in radio (synchrotron radiation from the hot spots), and UV, X- and gamma rays (IC radiation from the lobes), as
\begin{equation}
F\sim 0.1L_{\rm j}/4\pi r_{\rm c}^2(1+z)^2\approx 2\times 10^{-17}\left(L_{\rm j}/10^{44}{\rm erg~s}^{-1}\right)\,{\rm erg~s}^{-1}\,{\rm cm}^{-2},
\end{equation}
where $r_{\rm c}\approx 8427$~Mpc is the comoving radial distance to $z=6$. It has been assumed for simplicity that most of the lobe pressure is in the form of non-thermal electrons, and that $\sim 10$\% of the radiation energy goes to each specific energy band.

\subsection{Jet contribution to the reionization}\label{ji}

To assess the role of AGN jet lobes to contribute (maintaining and possibly furthering) the reionization of H at $z\sim 6-7$, we adopted the black-hole mass functions presented in Sect.~2, and fix for each black hole an ionizing luminosity $\propto L_{\rm j}\propto L_{\rm Edd}$. This yields a H-ionizing luminosity (comoving) density:
\begin{equation}
\dot\epsilon_{\rm ion}=\int\chi\,L_{\rm Edd}n_{\rm BH}(M_{\rm BH}){\rm d}M_{\rm BH}\,,
\end{equation}
where $n_{\rm BH}=\phi/M_{\rm BH}\ln(10)$ is the differential (comoving) density of the black-hole mass function \citep[see, e.g. Eq.~6 in][]{shankar09}, and the parameter $\chi$ includes departures from the optimal scenario, meaning: (i) a jetted AGN fraction of less than one; (ii) an AGN duty cycle $<100$\%; (iii) $L_{\rm j}<L_{\rm Edd}$; (iv) a non-dominant lobe non-thermal pressure; and (v) IC radiation mostly below or above UV \cite[in the latter case heating rather than ionizing the medium H, e.g.][]{shull85,mirabel11}. We adopted $\chi=0.001$ as a conservative case (A), and $\chi=0.1$ as a very optimistic case (B). Recall that under optimal conditions and $L_{\rm j}>L_{\rm Edd}$, $\chi>1$ is formally possible.
%Note that in the case of (i), (ii) and (iii), both low and high efficiencies are plausible; in the case of (iv) and (v), efficiencies of $\gtrsim 0.1$ seem reasonable. Finally, $\chi$ may also be larger than one if $L_{\rm j}>L_{\rm Edd}$, which is not unfeasible.

For simplicity, we compare $\dot\epsilon_{\rm ion}$ at $z\sim 6$ with the required H-ionizing luminosity (comoving) density of the whole reionization epoch, $\dot\epsilon_{\rm obs}$, estimated by \cite{madau17} as approximately three ionizing photons per H-atom and Gyr ($\dot\epsilon_{\rm obs}\approx 10^{40}$~erg~s$^{-1}$~Mpc$^{-3}$ in 13.6~eV photons). We note, however, that a more accurate comparison should properly account for $n_{\rm BH}$ at $z>6$, when a strong decline in the empirical $n_{\rm BH}$ is expected with $z$ \citep[e.g.][]{hopkins07,willott10}. 

The result of the comparison is the following: the computed $\dot\epsilon_{\rm ion}$ for cases A and B are $\approx 2\times 10^{37}$~erg~s$^{-1}$~Mpc$^{-3}$ (\citealt{willott10}; $\approx 10^{38}$~erg~s$^{-1}$~Mpc$^{-3}$ for \citealt{shankar09}) and $\dot\epsilon_{\rm ion}\approx 2\times 10^{39}$~erg~s$^{-1}$~Mpc$^{-3}$ (\citealt{willott10}; $\approx 10^{40}$~erg~s$^{-1}$~Mpc$^{-3}$ for \citealt{shankar09}), respectively. 
Even if very approximate, these results already indicate that jet lobes at $z\gtrsim 6$ may have produced $\lesssim 1$\% of the H-ionizing photons in a pessimistic scenario, but their contribution may have been non-negligible ($\gtrsim 20$\%) under more favorable conditions.

\section{Discussion}\label{dis}

As mentioned in the previous section, to compare with $\dot\epsilon_{\rm obs}$, $\dot\epsilon_{\rm ion}$ was assumed to be constant for the whole duration of the reionization epoch, but this was likely not the case. For $z>6-7$, in addition to the expected decline of the empirical black-hole mass function, the masses of the growing black holes, and thus their $L_{\rm j}$, are expected to be lower. On the other hand, the actual number of (jetted) black holes may have been larger than those empirically inferred. For instance, it has been proposed that active IMBH may have been present in all gas-rich dwarf galaxies at $z\gtrsim 6$, producing outflows and possibly jets (e.g. \citealt{silk17,barai18}; see however \citealt{latif18}). In fact, \cite{salvador17} predicts $\sim 100$ times more IMBH ($M\sim 10^5$~M$_\odot$) at redshift $z\sim 6$ than the empirical mass functions \citep{willott10}. If for instance 10\% of those IMBH untraced by empirical studies had produced jets with $L_{\rm j}\sim 0.1\,L_{\rm Edd}$, their contribution to $\dot\epsilon_{\rm ion}$ may have been relevant, also at $z>6-7$.

In addition to the AGN luminosity function, or the black-hole mass function, the accretion rate (comoving) density ($\dot\rho_{\rm acc}$) at $z\sim 6$ derived from X-ray background studies can set a limit on the ionizing role of AGN jet lobes assuming $\dot\epsilon_{\rm ion}<\dot\rho_{\rm acc}\,c^2$. \cite{vito18} estimates the accretion rate (comoving) density at $z\sim 6$ as $\dot\rho_{\rm acc}\sim 10^{-6}\,$M$_\odot$~yr$^{-1}$~Mpc$^{-3}$ \citep{vito18}, which yields $\dot\epsilon_{\rm ion}<6\times 10^{40}$~erg~s$^{-1}$~Mpc$^{-3}$, leaving room for $\dot\epsilon_{\rm ion}$ to significantly contribute to $\dot\epsilon_{\rm obs}$. The estimate for $\dot\rho_{\rm acc}$ is model dependent and possibly too conservative, but is not inconsistent with AGN jets being relevant H-ionizing sources at $z\gtrsim 6$. The black-hole mass (comoving) densities of the adopted mass functions are $\sim 10^2-10^3$~M$_\odot$, again consistent with the (model-dependent) limits found in the literature \citep[$\sim 10^3-10^4$~M$_\odot$; e.g.][]{hopkins07,salvaterra12,comastri15,cappelluti17}.

It is worth checking, in the jetted AGN scenario studied here, whether $\dot\epsilon_{\rm ion}$ could represent a non-negligible contribution to the reionization without overcoming the cosmic radiation background (CRB). For the black-hole mass functions adopted here, one can provide an estimate of the broadband flux as:
\begin{equation}
F_\Omega\sim 0.1\dot\epsilon_{\rm ion}V_{\rm c}/16\pi^2 r_{\rm c}^2(1+z)^2=
\end{equation}
$$
=0.1\dot\epsilon_{\rm ion}r_{\rm c}/12\pi(1+z)^2\approx (1-4)\times 10^{-4}(\chi/0.1)\,
{\rm nW}\,{\rm m}^{-2}\,{\rm srad}^{-1},
$$
where $V_{\rm c}=4\pi r_{\rm c}^3/3$ is the comoving volume of the Universe, $z=6$, and a constant source behaviour with $z$ has been assumed. The value of $F_\Omega$ is an order of magnitude estimate, but can be directly compared with the spectral energy distribution ($\lambda F^{\rm obs}_{\Omega\lambda}$) shown for example in Fig.~3 in \cite{cooray16}, which presents values ranging from $\sim 10^{-3}$ (radio, gamma rays) to $10^{-1}$~nW~m$^{-2}$~srad$^{-1}$ (X-rays). For instance, for an optimistic $\chi\sim 0.1$, the jetted AGN at $z\gtrsim 6$ may have had an important role ionizing the Universe without violating (but somewhat contributing to) the CRB.

Future, more accurate studies should be devoted to further explore the possibility that AGN jets may have contributed to the reionization of the IGM at $z\sim 6$ and earlier. In particular, more detailed population prescriptions, including yet untraced IMBH populations, an empirically derived accretion-jet relation for high-$z$ AGN, and a more precise model for the jet lobe dynamics and emission (including the shocked IGM thermal component), would give better constraints on the importance of AGN jets ionizing (and heating) the IGM at $z\gtrsim 6$. It would also be valuable to explore the impact of jet lobe relativistic electrons on the CMB spectrum through IC, as arcminute-scale small distortions may be expected \citep[e.g.][]{yamada99,malu17}.

\section{Summary}\label{sum}

In this work, we propose that, at $z\gtrsim 6$: (i) accretion was efficient \citep[following, e.g.][]{willott10,shankar13}; (ii) jetted AGN were common \citep[following, e.g.][]{sbarrato15}; (iii) jets carried a good fraction of the accretion energy \citep[e.g.][]{ghisellini14,sbarrato16}; and (iv) AGN jet-inflated lobe pressure has a significant non-thermal electron contribution \citep[e.g.][]{croston18}. In addition, we show that a substantial fraction of the jet power could have gone to ionize (and heat) the IGM at $z\gtrsim 6$ through CMB IC scattered photons in the lobes. Taking all this into account, and adopting empirical mass functions of black holes at $z\sim 6$ and extrapolating them to the late reionization epoch ($z\sim 6-7$) \citep[possibly a conservative characterization of the actual population, e.g.][]{salvador17,silk17,barai18}, we predict that AGN jet lobes may have contributed non-negligibly to the reionization of the Universe at $z\gtrsim 6$.

\begin{acknowledgements}
V.B-R. wants to thank an anonymous referee for constructive and useful comments that helped to improve the manuscript.
V.B-R. wants also to thank Kazushi Iwasawa, Jordi Miralda-Escud\'e, Massimo Ricotti, and Eduard Salvador for interesting and fruitful discussions on the early Universe. 
V.B-R. acknowledges support by the Spanish Ministerio de Econom\'{i}a y Competitividad (MINECO/FEDER, UE) under grant AYA2016-76012-C3-1-P, with partial support by the European Regional Development Fund (ERDF/FEDER), MDM-2014-0369 of ICCUB (Unidad de Excelencia `Mar\'{i}a de Maeztu'), and the Catalan DEC grant 2017 SGR 643. 
\end{acknowledgements}

\FloatBarrier
\bibliographystyle{aa}
\bibliography{refs}

\end{document}